\documentclass[aps,prb,preprint,notitlepage]{revtex4-1}
\usepackage{amsmath}
\usepackage{graphicx}
\usepackage{amssymb}

\usepackage{gt17}
\usepackage{color} 
\usepackage{graphicx}
\usepackage{bm}
\usepackage{amsmath}
\begin{document}
\title{
Optical responses induced by spin gauge field at the second order
}
\author{E. Karashtin}
\email{eugenk@ipmras.ru}
\affiliation{University of Nizhny Novgorod, 23 Prospekt Gagarina, 603950, Nizhny Novgorod, Russia}
\affiliation{Institute for Physics of Microstructures RAS, GSP-105, 603950, Nizhny Novgorod, Russia}
\author{Gen Tatara} 
\affiliation{RIKEN Center for Emergent Matter Science (CEMS), 
2-1 Hirosawa, Wako, Saitama, 351-0198 Japan}

\date{\today}

\begin{abstract}
Optical responses of ferromagnetic materials with spin gauge field that drives intrinsic spin curren is theoretically studied.
The conductivity tensor is calculated based on a linear response theory to the applied electric field taking account of the non-linear effects of the spin gauge field to the second order.
We consider the case where the spin gauge field is uniform, realized for spiral magnetization structure or uniform spin-orbit interaction.
The spin gauge field, or an intrinsic spin current, turns out to give rise to anisotropic optical responses, which is expected to be useful to experimental detection of  magnetization structures.  
\end{abstract}  

\maketitle

\section{Introduction}
Spin current plays central roles in spintronics. 
Moreover, intrinsic spin current in solids has been shown to induce several peculiar interactions. 
As spin current breaks inversion symmetry keeping the time reversal symmetry, the interactions induced by spin current have the same symmetry \cite{TataraReview19}, like  the antisymmetric Dzyaloshinskii-Moriya (DM) interaction between spins.
It was theoretically shown that the DM interaction constant is $D_i^a=\hbar a^3 j_{{\rm s},i}^a$, where $i$ and $a$ represent directions in space and spin, respectively, and $a$ is the lattice constant. Namely, it is proportional to the expectation value of the spin current $j_{{\rm s},i}^a$ intrinsically induced by the spin-orbit interaction when inversion symmetry is broken \cite{Kikuchi16}. 
In this picture, the origin of the DM interaction is the Doppler shift due to the intrinsic flow of spin. 
The picture naturally explains the Doppler shift of spin waves in the presence of the DM interaction observed in Refs. \cite{Iguchi15,Seki16}. 
The spin current picture was explored further in Refs. \cite{FreimuthDM17,Koretsune18}.
The relation between the DM constant and spin current has been experimentally confirmed by injecting external spin current \cite{Karnad18,Kato19}.
In the case of insulator magnets, Ref. \cite{Katsura05} pointed out that an electric polarization is induced by the vector chirality, which is equivalent to an intrinsic spin current. 
Similar  flexo-magnetoelectric effects due to intrinsic spin current were discussed in Refs. \cite{62Bruno,Zvezdin12}.
The second harmonic generation was discussed to arise from intrinsic spin current under the action of optical electromagnetic wave \cite{Wang1,Werake1,Karashtin1,Karashtin3}.

The Doppler shift picture can be applied for discussing anomalous optical properties of solids.
When a spin-orbit interaction breaking inversion symmetry, like the Rashba spin-orbit interaction, coexists with interactions breaking time-reversal invariance, like the coupling to a magnetization, intrinsic charge flow is allowed, resulting in a directional dichroism of light \cite{Shibata16,Shibata18}.
The dichroism here is induced by the effective gauge field proportional to $\uv\equiv \bm{\alpha}_{\rm R}\times\bm{M}$, where  $\bm{\alpha}_{\rm R}$ and $\bm{M}$ denote the Rashba field and the magnetization, respectively.
The effective gauge field induces charge current, and the effect is viewed as Doppler shift of light, mentioned in Ref. \cite{Sawada05}.
It was in fact demonstrated theoretically that the Rashba interaction and magnetization leads to a coupling proportional to $\uv\cdot(\Ev\times\Bv)$ between the electric and magnetic fields, $\Ev$ and $\Bv$ \cite{Kawaguchi16}.
This coupling is equivalent to switching to a moving frame with velocity $\uv$.
As seen above,  various intrinsic flow or effective gauge field causing Doppler shift in solids have been detected as asymmetric transport effects \cite{TataraReview19}.

Spin current is manipulated by spin gauge fields  \cite{TataraReview19}, and spin-orbit interaction argued above is an example of spin gauge field driving the spin current. 
Slowly-varying spin textures also act as spin gauge field for conduction electron \cite{Volovik87,TKS_PR08}.
In this paper, we study nonlinear effects of intrinsic spin currents induced by spin gauge field on the optical properties of metallic systems. 
We include the spin gauge field to the second order, and derive linear response expression of the optical conductivity. 
We consider the case of uniform spin current, i.e., spin gauge field uniform in space and time. 
We have in mind the case where one spin gauge field is due to an intrinsic spin-orbit interaction like Rashba interaction and another generated by magnetization structures. 
As magnetization structure, we consider a spiral spin structure where the spins change with a constant pitch.

\section{Formalism}
The spin gauge field approach for spin structures is valid in the adiabatic regime where the electron spin follows the magnetization structure due to $sd$ exchange interaction \cite{TataraReview19}. 
The present manuscript therefore studies a strong $sd$ exchange case, different from the perturbative treatment of the $sd$ exchange interaction carried out in Ref. \cite{Shibata16}.
The model we consider is the $sd$ exchange model described by a Hamiltonian
\begin{align}
 H&=\intr c^\dagger \lt(-\frac{\nabla^2}{2m}-\ef-M\nv(\rv)\cdot\sigmav \rt)c
\end{align}
where $c$ and $c^\dagger$ are electron field operators, $\ef$ is the Fermi energy, $m$ is the electron mass, $M$ is the spin splitting energy due to the $sd$ exchange interaction, and $\nv(\rv)$ is a unit vector field representing the magnetization direction, $\sigmav$ being a vector or Pauli matrices.
We apply a unitary transformation in the spin space to diagonalize the exchange interaction \cite{TKS_PR08}. 
The new electron field in the transformed rotated frame is $\tilde{c}\equiv U^{-1}(\rv)c$, where $U(\rv)$ is a $2\times 2$ unitary matrix, which is chosen to satisfy $U^{-1}(\nv(\rv)\cdot\sigmav)U=\sigma_z$.
The Hamiltonian in the rotated frame reads 
\begin{align}
 H&=\intr \tilde{c}^\dagger \lt(-\frac{1}{2m}\lt(\nabla_i+iA_{{\rm s},i}\rt)^2-\ef-M\sigma_z \rt)\tilde{c}
\end{align}
where $A_{{\rm s},i}\equiv -iU^{-1}\nabla_i U$ is the spin gauge field of a $2\times2$ matrix.

For the uniform spin gauge fields, the conductivity tensor at linear response to the applied electric field with angular frequency $\Omega$ can be calculated straightfowardly including all the orders of the spin gauge field because the spin gauge field does not carry wave vectors nor angular frequency. Nevertheless we focus on the second order effects at the end.  
We denote in this section the uniform total spin gauge field as $A_{{\rm s},i}^\alpha$, where $i$ and $\alpha$ are directions in space and spin space, respectively, and derive the expression for the conductivity tensor.
Later  in the next section various spin gauge fields are introduced, and the results in this section should be read replacing $A_{{\rm s},i}^\alpha$ by the sum of all the spin gauge fields of interest.  
The conductivity tensor is 
\begin{align}
 \sigma_{ij}(\Omega)&=\frac{1}{-i\Omega}\sumom\sumkv 
 \tr[v_i G_{\kv\omega}v_j G_{\kv,\omega+\Omega}+\delta_{ij}G_{\kv\omega}]^<
 \label{sigmadef}
\end{align}
Here  the velocity operator is
\begin{align}
 v_i &= \frac{k_i}{m}+A_{{\rm s},i}^\alpha \sigma_\alpha
\end{align}
and the Green's function includes spin gauge field $A_{\rm s}$ and $^<$ denotes the lesser component.
The last term  of Eq. (\ref{sigmadef}) arise from the 'diamagnetic' contribution to the electric current, namely the second term of $j_i=v_i+A_i$, where $A$ represents the gauge field of electromagnitism.
Retarded Green's function is 
\begin{align}
G^\ret_{\kv\omega} &= \frac{1}{\omega-\frac{k^2}{2m}-\sum_{i\alpha}
(\gammav_k)_\alpha\sigma_\alpha +i\eta}
\end{align}
where 
\begin{align}
(\gammav_k)_\alpha & \equiv  M_\alpha+\sum_i k_i  A_{{\rm s},i}^\alpha
 \end{align}
$\eta$ represents small positive imaginary part due to the electron damping, and 
$\Mv\equiv M\sigma_z$ is the diagonalized spin splitting. 
(Being in the rotated frame, $\Mv$ is diagonalized along the $z$ axis.) 
Evaluating the lesser component, we have 
\begin{align}
 \sigma_{ij}(\Omega)
 &= \frac{1}{-i\Omega}\sumom\sumkv \nnr
 & \tr\biggl[ 
 (f(\omega+\Omega)-f(\omega)) v_i G^\ret_{\kv\omega}v_j G^\adv_{\kv,\omega+\Omega}
  +f(\omega) v_i G^\adv_{\kv\omega}v_j G^\adv_{\kv,\omega+\Omega}
  -f(\omega+\Omega) v_i G^\ret_{\kv\omega}v_j G^\ret_{\kv,\omega+\Omega}
  \nnr
  &  +\delta_{ij}f(\omega) \lt(  G^\adv_{\kv\omega}-G^\ret_{\kv\omega}\rt) 
\biggr]
\label{sigmaijra}
\end{align}
where $f(\omega)\equiv [e^{\beta \omega}+1]^{-1}$ is the Fermi distribution function ($\beta=(\kb T)^{-1}$ is the inverse temperature).
Trace of spin ($a,b=\ret,\adv$),
\begin{align}
\sumkv \tr[v_i G^a_{\kv\omega}v_j G^b_{\kv,\omega'}] 
& \equiv \sumkv  K^{ab}_{ij} (\kv,\omega,\omega')
\end{align} 
is written defining  
\begin{align}
G^a_{\kv\omega}=\frac{\pi_{k\omega}^a+\gammav_k\cdot\sigma}{\Pi_{k\omega}^a} 
\end{align}
where ($\ekv=\frac{k^2}{2m}-\ef$) 
\begin{align}
\pi_{k\omega}^\ret & \equiv \omega-\ekv+i\eta , &
\Pi_{k\omega}^a  \equiv (\pi_{k\omega}^a)^2 -\gamma_k^2 , 
\end{align}
as
\begin{align}
K^{ab}_{ij}&(\kv,\omega,\omega')  =
\frac{1}{ \Pi_{k\omega}^a\Pi_{k,\omega'}^b } \nnr
& \times 
\tr \biggl[ \lt(\frac{k_i}{m}+{\Av}_{{\rm s},i}\cdot\sigmav\rt) \lt( \pi_{k\omega}^a+(\Mv+k_k {\Av}_{{\rm s},k})\cdot\sigmav \rt)  
\lt(\frac{k_j}{m}+{\Av}_{{\rm s},j}\cdot\sigmav\rt)
\lt( \pi_{k,\omega'}^b+(\Mv+k_l {\Av}_{{\rm s},l})\cdot\sigmav \rt) \biggr]
\end{align}
Focusing on the second order contribution in the spin gauge field, we obtain 

\begin{align}
\sumkv & K^{ab}_{ij}(\kv,\omega,\omega') 
 = 2\biggl[
\frac{1}{m^2} (\kappa_{ij}^{(2)ab}+M^2\kappa_{ij}^{(0)ab}) 
+ \frac{1}{m} \lt( \kappa_i^{ab} ({\Av}_{{\rm s},j}\cdot\Mv) +\kappa_j^{ab} ({\Av}_{{\rm s},i}\cdot\Mv) \rt) \nnr
& + \frac{1}{m}(\kappa_{ik}^{(1)ab}({\Av}_{{\rm s},j}\cdot{\Av}_{{\rm s},k})+\kappa_{jk}^{(1)ab}({\Av}_{{\rm s},i}\cdot{\Av}_{{\rm s},k}) )
    + \frac{2}{m^2}\kappa_{ijk}^{ab} ({\Av}_{{\rm s},k}\cdot\Mv) +
     \frac{1}{m^2}\kappa_{ijkl}^{ab} ({\Av}_{{\rm s},k}\cdot{\Av}_{{\rm s},l})\nnr
       &
+\lt[ \lambda^{(2)ab} -M^2 \lambda^{(0)ab} \rt] ({\Av}_{{\rm s},i}\cdot{\Av}_{{\rm s},j})
+ 2\lambda^{(0)ab}({\Av}_{{\rm s},i}\cdot\Mv)({\Av}_{{\rm s},j}\cdot\Mv)
     - i  \Omega \lambda^{(0)ab}({\Av}_{{\rm s},i}\times {\Av}_{{\rm s},j})\cdot\Mv\biggr]
  \label{Kijcalculation2}
\end{align}
where the coefficients are 
\begin{align}
\sumkv  \frac{1}{ \Pi_{k\omega}^a\Pi_{k,\omega'}^b }  
& \equiv \lambda^{(0)ab}(\omega,\omega') , 
\sumkv  \frac{\pi_{k\omega}^a\pi_{k\omega'}^b}{ \Pi_{k\omega}^a\Pi_{k,\omega'}^b }  
 \equiv \lambda^{(2)ab}(\omega,\omega')  \nnr
\sumkv  \frac{k_ik_j}{ \Pi_{k\omega}^a\Pi_{k,\omega'}^b }  
& \equiv \kappa_{ij}^{(0)ab}(\omega,\omega') ,  \sumkv  k_ik_j \frac{\pi_{k\omega}^a+\pi_{k\omega'}^b}{ \Pi_{k\omega}^a\Pi_{k,\omega'}^b }  
 \equiv \kappa_{ij}^{(1)ab}(\omega,\omega') , \sumkv k_ik_j \frac{\pi_{k\omega}^a\pi_{k\omega'}^b}{ \Pi_{k\omega}^a\Pi_{k,\omega'}^b }  
 \equiv \kappa_{ij}^{(2)ab}(\omega,\omega')  \nnr
 \sumkv k_i \frac{\pi_{k\omega}^a+\pi_{k\omega'}^b}{ \Pi_{k\omega}^a\Pi_{k,\omega'}^b }  
& \equiv \kappa_{i}^{ab}(\omega,\omega')  , 
 \sumkv k_ik_jk_k \frac{1}{ \Pi_{k\omega}^a\Pi_{k,\omega'}^b }  
 \equiv \kappa_{ijk}^{ab}(\omega,\omega')  ,
 \sumkv k_ik_jk_k k_l \frac{1}{ \Pi_{k\omega}^a\Pi_{k,\omega'}^b }  
 \equiv \kappa_{ijkl}^{ab}(\omega,\omega')  
\end{align}
Note that contributions odd in $\kv$ such as $ \kappa_{i}^{ab}$ and $\kappa_{ijk}^{ab}$ are finite because the energy $\ekv\pm\gamma_k$ contains contribution odd in $\kv$ when $M\neq0$. 
The summations over  $\kv$ are evalulated expanding $\gamma_k$ in $\Pi_{k\omega}^a$ with respect to the spin gauge field.
The odd terms turn out to be 
\begin{align}
 \kappa_{i}^{ab}(\omega,\omega')  
   &=  \kappa_{1}^{ab}(\omega,\omega') ( {\Av}_{{\rm s},i} \cdot\Mv) \nnr
 \kappa_{ijk}^{ab}(\omega,\omega')  
    &=\kappa_3^{ab}(\delta_{ij} {\Av}_{{\rm s},k}+\delta_{ik} {\Av}_{{\rm s},j}+\delta_{jk} {\Av}_{{\rm s},i})\cdot \Mv 
\end{align}
where $\kappa_{1}^{ab}$ and $\kappa_{3}^{ab}$ do not depend on the spin gauge field to the lowest order.
Other coefficients depend on the gauge field to the lowest order as ($\mu=0,1,2$)
\begin{align}
\lambda^{(\mu)ab} 
  &= \lambda^{(\mu)ab}_{0}+  \lambda^{(\mu)ab}_{2}\sum_{k}({\Av}_{{\rm s},k}\cdot\Mv)^2 \nnr
\kappa_{ij}^{(\mu)ab} 
   &= \delta_{ij} [ \kappa_{0}^{(\mu)ab} +\kappa_{{\rm d}2}^{(\mu)ab} \sum_{k}({\Av}_{{\rm s},k}\cdot\Mv)^2 ]
       +\kappa_{2}^{(\mu)ab}  ({\Av}_{{\rm s},i} \cdot \Mv)({\Av}_{{\rm s},j}\cdot\Mv)   \nnr
\kappa_{ijkl}^{ab}
   &= \kappa_4^{ab}(\delta_{ij}\delta_{kl}+\delta_{ik}\delta_{jl} +\delta_{il}\delta_{jk}) 
\end{align}
where ${\rm d}$ denotes diagonal and  $\lambda^{(\mu)ab}_{\nu}$, $\kappa_{\nu}^{(\mu)ab} $ ($\nu=0,2, {\rm d}2$) and $ \kappa_4^{ab}$ do not depend on the gauge field to the lowest order.

\section{Result}
From the above consideration, the conductivity tensor to the second order in the spin gauge field is written as 
\begin{align}
\sigma_{ij}
&=  \delta_{ij}[\chi_0+\chi_{{\rm d}2} \sum_{k} ({\Av}_{{\rm s},k}\cdot{\Av}_{{\rm s},k})+\chi_{{\rm d}2}^{\rm (ad)}  \sum_{k} ({\Av}_{{\rm s},k}\cdot\Mv)({\Av}_{{\rm s},k}\cdot\Mv)] \nnr
& +\chi_2({\Av}_{{\rm s},i}\cdot{\Av}_{{\rm s},j})
+\chi_{2}^{\rm (ad)} ({\Av}_{{\rm s},i}\cdot\Mv)({\Av}_{{\rm s},j}\cdot\Mv)
+\chi_3 ({\Av}_{{\rm s},i}\times {\Av}_{{\rm s},j})\cdot\Mv
\label{sigmafinalresult}
\end{align}
where $\chi_i$'s are functions of the external angular frequency $\Omega$.
The scalar product $({\Av}_{{\rm s},i}\cdot\Mv)$ represents the adiabatic component 
(denoted by $^{\rm (ad)}$) of the spin gauge field, i.e., the component along the magnetization.
The nonadiabatic (perpendicular) components of the gauge field affects the terms with $\chi_{{\rm d}2}$, $\chi_2$ and $\chi_3$.  

Although the form in Eq. (\ref{sigmafinalresult}) is natural from the symmetry consideration, the expressions for the coefficients are in principle known by the present microscopic study.
For example, we have 
\begin{align}
 \chi_2(\Omega) &= \frac{1}{-i\Omega} \sumom \biggl[ 
 (f(\omega+\Omega)-f(\omega)) \tilde{\chi_2}^{\ret\adv}(\omega,\omega+\Omega) 
  +f(\omega)  \tilde{\chi_2}^{\adv\adv}(\omega,\omega+\Omega)
  -f(\omega+\Omega) \tilde{\chi_2}^{\ret\ret}(\omega,\omega+\Omega) 
\biggr]  \nnr 
\chi_3(\Omega) &= \sumom \biggl[ 
 (f(\omega+\Omega)-f(\omega)) \lambda^{(0)\ret\adv}(\omega,\omega+\Omega) 
  +f(\omega)  \lambda^{(0)\adv\adv}(\omega,\omega+\Omega)
  -f(\omega+\Omega) \lambda^{(0)\ret\ret}(\omega,\omega+\Omega) 
\biggr]  
\label{chi2chi3def}
\end{align}
where 
\begin{align}
 \tilde{\chi_2}^{ab}(\omega,\omega') 
  &\equiv 2\lt[
    \lambda^{(2)}(\omega,\omega')-M^2 \lambda^{(0)}(\omega,\omega') +\frac{2}{m}\kappa_{0}^{(1)}(\omega,\omega')
    +\frac{2}{m^2} \kappa_4(\omega,\omega') \rt]^{ab}
\end{align}

The coefficients $\chi_i$'s  are finite at $\Omega=0$, in spite of the factor of $\Omega^{-1}$ in the definition, Eqs.  (\ref{sigmaijra})(\ref{chi2chi3def}). This is checked easily based on Eq. (\ref{sigmaijra}).
In fact, the square bracket in Eq. (\ref{sigmaijra}) vanishes linearly at $\Omega\ra0$, as 
$\sumkv \tr[v_i G_{\kv\omega}^\adv v_j G_{\kv\omega}^\adv]
=\sumkv \tr[(G_{\kv\omega}^\adv)^{-1}(\partial_{k_i} G_{\kv\omega}^\adv) v_j G_{\kv\omega}^\adv]
= -\delta_{ij}\sumkv \tr[G_{\kv\omega}^\adv]$, where we used 
$(\partial_{k_i} G_{\kv\omega}^\adv)=  G_{\kv\omega}^\adv v_i  G_{\kv\omega}^\adv$ and integral by parts with respect to $\kv$.

In the low frequency limit ($\Omega\ra0$), the expression of the conductivity Eq. (\ref{sigmaijra}) is simplified to be 
\begin{align}
 \sigma_{ij}(\Omega\ra0)
 & = \sumom f'(\omega)  \sumkv \lt[ 
 K^{\ret\adv}_{ij} (\kv,\omega,\omega)
  -\frac{1}{2}\lt(  K^{\adv\adv}_{ij} (\kv,\omega,\omega)+ K^{\ret\ret}_{ij} (\kv,\omega,\omega) \rt)\rt]
  \nnr
  &\simeq - \frac{1}{2\pi} \sumkv \lt[ 
  K^{\ret\adv}_{ij} (\kv,0,0) 
  -\frac{1}{2}\lt( K^{\adv\adv}_{ij} (\kv,0,0)+ K^{\ret\ret}_{ij} (\kv,0,0) \rt)\rt]
 \end{align}
 where we used $f'(\omega)\simeq - \delta(\omega)$ assuming low temperatures in the last line.

 The parameter $\chi_2$ in Eq,. (\ref{sigmafinalresult}) characterizes the magnitude of anisotropy of optical response. 
 Let us derive explicit expression for $\chi_2$.
 Using 
 \begin{align}
  \frac{1}{ \Pi_{k\omega}^a } =\frac{1}{\gamma_\kv}\sum_{\sigma=\pm} \sigma g^a_{\kv\omega\sigma}  
 \end{align}
where
 \begin{align}
 g^a_{\kv\omega\sigma} &\equiv \frac{1}{ \pi_{k\omega}^a -\sigma \gamma_{\kv}}
 \end{align}
is the spin-polarized Green's function,
we obtain 
\begin{align}
 \lambda^{(2)ab}(\omega,\omega')
 &= \frac{1}{4}\sum_{\kv\sigma}(g^a_{\kv\omega\sigma} g^b_{\kv\omega'\sigma}+g^a_{\kv\omega\sigma}g^b_{\kv\omega',-\sigma}) \nnr
 \lambda^{(0)ab}(\omega,\omega')
 &= \frac{1}{4}\sum_{\kv\sigma}\frac{1}{(\gamma_k)^2} g^a_{\kv\omega\sigma}( g^b_{\kv\omega'\sigma}-g^b_{\kv\omega',-\sigma}) \nnr
 \kappa_{0}^{(1)ab}(\omega,\omega') 
 &= \frac{1}{12}\sum_{\kv\sigma}\frac{k^2}{\gamma_k} \sigma g^a_{\kv\omega\sigma} g^b_{\kv\omega'\sigma} \nnr
 \kappa_4^{ab}(\omega,\omega') 
 &= \frac{1}{60}\sum_{\kv\sigma}\frac{k^4}{(\gamma_k)^2} g^a_{\kv\omega\sigma}( g^b_{\kv\omega'\sigma}-g^b_{\kv\omega',-\sigma}) 
\end{align}
Those coefficients with $a=\ret$ and $b=\adv$ turn out to be dominant for $\eta/M\ll1$.
Moreover, contributions containing the Green's functions with different spins are neglected for $\eta/M\ll1$.
Considering low frequency ($\Omega\ra0$) limit, the coefficient $\chi_2$ is 
\begin{align}
 \chi_2(\Omega) &= -i  \sum_{\sigma}\nu_\sigma \tau_\sigma\lt(1+\frac{(k_\sigma)^2}{3mM}+\frac{(k_\sigma)^4}{15m^2M^2}\rt)
 \label{chi2_2}
\end{align}
where $\nu_\sigma$, $k_\sigma$ and $\tau_\sigma(\equiv 1/(2\eta_\sigma))$ are spin-resolved electron density of states, Fermi wave vector and elastic lifetime, respectively.
Noting that the Boltzmann conductivity is $\sigma_0\sim \sum_\sigma \nu_\sigma (k_\sigma)^2 \tau_\sigma$,  the anisotropic terms induced by spin gauge field is of the relative order of 
$(A_{\rm s}/\kf)^2$ ($\kf$ being the Fermi wave vector) compared to $\sigma_0$.
For spiral magentization structure with a pitch $Q$, this ratio is $A_{\rm s}/\kf \sim (Q/\kf)$ and for Rashba spin gauge field, 
$A_{\rm R}/\kf \sim \alpha_{\rm R}(\kf)^2/\ef$, as will be discussed in the next section. 
For a short spiral wave length (several nanometers like in Ho \cite{Koehler65}) and for large Rashba coupling like in BiTeI \cite{Ishizaka11}, the anisotropy would be easily detected experimentally.

 \section{Application to spiral structures}
\begin{figure}
 \includegraphics[width=0.4\hsize]{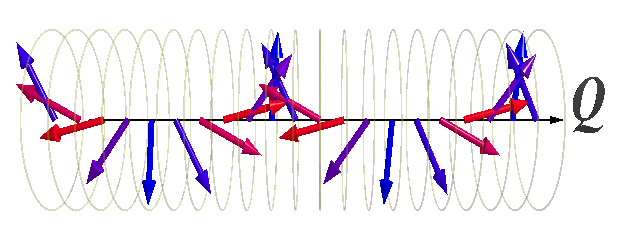}
 \includegraphics[width=0.4\hsize]{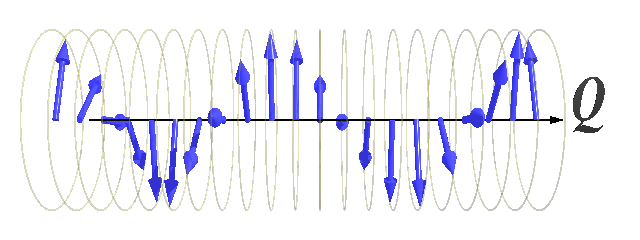}
 \caption{Magnetization structure of two spirals, N\'eel (left) and Bloch (right). The direction of the spiral denoted by $\bm{Q}$ is the $x$-axis.\label{FIGspiral}} 
\end{figure}

%
We consider examples of spiral magnetization structures (Fig. \ref{FIGspiral}).

\subsection{N\'eel type spiral}
The first one is a N\'eel type spiral along the $x$-direction;
\begin{align}
  \nv(\rv)&= \xvhat \sin Qx+\zvhat \cos Qx \label{Neelspiral}
\end{align}
where $\hat{\ }$ denotes the unit vector along the coordinate axis and $Q$ is the pitch of the spiral.
Equilibrium spin current induced by magnetization textures is given by \cite{TataraReview19}  
$\jv_{{\rm s},i}=\nv\times\nabla_i\nv$, which in the present case is perpendicular to the magnetization plane ($xz$-plane); 
$\jv_{{\rm s},i}=\delta_{i,x}Q\yvhat$.
The unitary transformation to diagonalize the $sd$ exchange interaction for this magnetization structure is $U=\mv\cdot\sigmav$, where 
\begin{align}
 \mv=\lt(\sin\frac{Qx}{2},0,\cos\frac{Qx}{2} \rt)
\end{align}
The spin gauge field arising from the structure is $A_{{\rm s},i}^\alpha=\frac{-i}{2}\tr[\sigma_\alpha U^{-1}\nabla_i U]$, which for the N\'eel structure is
\begin{align}
 A_{{\rm s},i}^{{\rm N},\alpha} &= \frac{Q}{2}\delta_{i,x} \delta_{\alpha,y}
 \label{AsN}
\end{align}
The direction of the spin polarization of the gauge field is $y$, which is consistent with the equilibrium spin current flow.

\subsection{Bloch type spiral}
The second one is a bloch type spiral along the $x$-direction, where magnetization rotates in the plane perpendicular to the $x$-direction,
\begin{align}
  \nv(\rv)&= \yvhat \sin Qx+\zvhat \cos Qx  \label{Blochspiral}
\end{align}
The vector $\mv$ is 
\begin{align}
 \mv=\lt(0,\sin\frac{Qx}{2},\cos\frac{Qx}{2} \rt)
\end{align}
and the equilibrium spin current is  
$\jv_{{\rm s},i}=-\delta_{i,x}Q\xvhat$, and the spin gauge field arising from the structure is
\begin{align}
 A_{{\rm s},i}^{{\rm B},\alpha} &= - \frac{Q}{2}\delta_{i,x} \delta_{\alpha,x}
 \label{AsB}
\end{align}

Both N\'eel and Bloch type of spiral, the spin gauge field is finite for spatial direction of the spiral, i.e.,  $x$-axis, and so the second-order contribution to the conductivity tensor (Eq. (\ref{sigmafinalresult})) has only the diagonal components.
If the direction of the spiral deviates from the coordinate axis, symmetric off-diagonal components 
$\sigma_{ij}=\sigma_{ji}\propto  \Av_{{\rm s},i}\cdot \Av_{{\rm s},j}$ appear, where $ij$ denotes the directions in the plane containing the spiral direction.
The optical response can thus detect the direction of the intrinsic spin current induced by spin gauge field.

However Bloch and N\'eel spirals cannot be distinguished by the present optical response. 
This is because the optical response does not see the spin polarization direction (denoted by $\alpha$ of $A_{{\rm s},i}^\alpha$) but only the scalar product or the trace in the spin index (Eq. (\ref{sigmafinalresult})).
Spin direction affects optical response if an additional spin polarization is introduced by an external field or a spin-orbit interaction, which we consider in next two subsections.

\subsection{Spirals in an external magnetic field}

We consider the case of a magnetic field applied along the $x$-axis for Bloch and N\'eel spirals.
We  simply assume that the magnetization structure has a constant component of magnitude $\beta$ along the field without deriving solutions including magnetic field, and so the argument may not be applicable for large $\beta$. 
The magnetization profiles with $\beta$ are 
\begin{align} \label{Neelfield}
\nv_N &= \xvhat \frac{\beta + \sin Qx}{\sqrt{1 + \beta^2 + 2 \beta \sin Qx}} + \zvhat \frac{\cos Qx}{\sqrt{1 + \beta^2 + 2 \beta \sin Qx}},
\end{align}
\begin{align} \label{Blochfield}
\nv_B &= \xvhat \frac{\beta}{\sqrt{1 + \beta^2}} + \yvhat \frac{\sin Qx}{\sqrt{1 + \beta^2}} + \zvhat \frac{\cos Qx}{\sqrt{1 + \beta^2}}.
\end{align}
The tilted  Bloch case is the one representing the excitation around a magnetic skyrmion lattice \cite{Petrova11,Tatara14}.
Note that the limit $\beta \to 0$ corresponds to (\ref{Neelspiral}), (\ref{Blochspiral}), while $\beta \to \infty$ stands for a uniformly magnetized medium.
The vector $\mv$ which generates the unitary transformation to diagonalize the exchange interaction now has the form
\begin{equation} \label{mNeel}
\mv_N = \left(
\sin \frac{Q x}{2} \sqrt{\frac{1 - \frac{\cos Qx}{\sqrt{1 + \beta^2 + 2 \beta \sin Qx}}}{1 - \cos Qx}},0,
\cos \frac{Q x}{2} \sqrt{\frac{1 + \frac{\cos Qx}{\sqrt{1 + \beta^2 + 2 \beta \sin Qx}}}{1 + \cos Qx}} \right),
\end{equation}
\begin{align} \label{mBloch}
\bm{m}_B = \left( \beta \sin \frac{Qx}{2} \sqrt{\frac{1-\frac{\cos Qx}{\sqrt{1+\beta^2}}}{\left(1 - \cos Qx\right) \left(\beta^2 + \sin^2 Qx\right)}},
\sin \frac{Qx}{2}\sqrt{\frac{\left(1-\frac{\cos Qx}{\sqrt{1+\beta^2}}\right) \sin^2 Qx}{2 \left(\beta^2 + \sin^2 Qx\right)}},
\cos \frac{Qx}{2}\sqrt{\frac{1+\frac{\cos Qx}{\sqrt{1+\beta^2}}}{1 + \cos Qx}} \right)
\end{align}
for N\'eel and Bloch magnetization. 
 
In the N\'eel case, the vector-potential takes the form
\begin{align} \label{NeelA}
A_{{\rm s},i}^{{\rm N},\alpha} = \delta_{i,x} \delta_{\alpha,y}
\frac{Q}{2} \frac{\sin Q x}{\beta + \sin Q x} \sqrt{\frac{\left(\beta + \sin Q x\right)^2}{\sin^2 Q x}} \frac{1 + \beta \sin Q x}{1 + \beta^2 + 2 \beta \sin Q x}.
\end{align}
Obviously, it has the same component as it was with no mangetic field applied (see (\ref{AsN})). After averaging over the coordinates we arrive at
\begin{align} \label{NeelAaver}
\left<A_{{\rm s},i}^{{\rm N},\alpha}\right> = \left\{
\begin{matrix}
\delta_{i,x} \delta_{\alpha,y} \frac{Q}{2} \left( 1 - \frac{2}{\pi} \arctan \left|\beta\right|\right), \, \left|\beta\right| < 1 \\
\delta_{i,x} \delta_{\alpha,y} \frac{Q}{2} \frac{1}{\pi} \arcsin \frac{2 \left|\beta\right|}{1 + \beta^2}, \, \left|\beta\right| > 1.
\end{matrix}
\right.
\end{align}
where $\left<\ \right>$ stands for spatial averaging. This function continuously decreases from $\frac{Q}{2}$ to zero as $\beta$ grows.

Gauge field component having a perpendicular spin emerges if we apply an out-of plane  magnetic field   
for the Bloch case (\ref{mBloch}):
\begin{align} \label{BlochA} 
A_{{\rm s},i}^{{\rm B},\alpha} = \delta_{i,x} \frac{Q}{2} \left(
- \frac{\beta^2 \cos Qx + \sqrt{1+\beta^2} \sin^2 Qx}{\sqrt{1+\beta^2} \left(\beta^2 + \sin^2 Qx\right)} , 
\frac{\beta}{\sqrt{1+\beta^2}} \frac{\sqrt{\sin^2 Q x}}{1 + \frac{\cos Q x}{\sqrt{1+\beta^2}}}, 
\beta \frac{\tan \frac{Q x}{2} \cos Q x \left(1 - \frac{\cos Q x}{\sqrt{1 + \beta^2}}\right)}{\sqrt{\tan^2 \frac{Q x}{2}} \left(\beta^2 + \sin^2 Q x\right)}
\right).
\end{align}
After averaging over coordinates, $\left<A^{{\rm B},z}_{{\rm s} ,x} \right>= 0$, while two other 
 components are finite:
\begin{align} \label{BlochAaverX}
\left<A_{{\rm s},x}^{{\rm B},x}\right> = -\frac{Q}{2} \left(1 - \frac{\left|\beta\right|}{\sqrt{1+\beta^2}}\right),
\end{align}
\begin{align} \label{BlochAaverXz}
\left<A_{{\rm s},x}^{{\rm B},y}\right> = \frac{Q}{2} \frac{\beta}{\pi \sqrt{1 + \beta^2}} \log \frac{\sqrt{1 + \beta^2} + 1}{\sqrt{1 + \beta^2} - 1}.
\end{align}
One can see that (\ref{BlochAaverX}) is the same as (\ref{AsB}) that decays with the magnitude of magnetic oscillations. The $y$-component $\left<A_{{\rm s},x}^{{\rm B},y}\right>$ is odd with respect to $\beta$. It is zero at $\beta = 0 $ but has an infinite derivative at this point, thus growing very fast at small applied field. It reaches its maximum at $\beta^* \approx 0.66$ with the value $\left<A_{{\rm s},x}^{{\rm B},y}\right> \left(\beta^*\right) \approx 0.4  \left<A_{{\rm s},x}^{{\rm B},x}\right> \left(\beta = 0\right)$.

As we saw, response of the gauge field to a magnetic field depends much on the magnetic structure. 
Observation of optical response with an applied field is therefore expected to be useful to distinguish the structure.


\subsection{Spin-orbit interaction}
Let us consider spin-orbit interactions that break the inversion symmetry.
The first one is the Rashba interaction, whose Hamiltonian is 
\begin{align}
 H_{\rm R} &= -\frac{i}{2} \intr c^\dagger \alphav_{\rm R}\cdot(\nablalr\times\sigmav) c  
\end{align}
We consider first the case the Rashba field vector $\alphav_{\rm R}$ is along the $z$-axis.
In the rotated frame, the interaction reads 
\begin{align}
 H_{\rm R} &= -\frac{i}{2} \intr \tilde{c}^\dagger \alphav_{\rm R}\cdot(\nablalr\times\tilde{\sigmav}) \tilde{c} + \mbox{\rm spin density part}
\end{align}
where the first terms is the interaction between the spin current and the Rashba spin gauge field,  while the last term describing the spin density is neglected.
The electron field in the rotated frame is $ \tilde{c}\equiv U^{-1} c$ and $\tilde{\sigmav}=U^{-1} \sigmav U$ is the spin operator in the rotated frame.
The Rashba spin gauge field read from the interaction is 
\begin{align}
 A_{{\rm R},i} &= -im\epsilon_{ijk}\alpha_{{\rm R},j}\tilde{\sigma}_k \label{Rashbagf}
\end{align}

Explicit form for each magnetization profile is calculated using $\tilde{\sigma}_k=2m_k(\mv\cdot\sigmav)-\sigma_k$.
For the N\'eel type spiral, Eq. (\ref{Neelspiral}),  
\begin{align} \label{tildesigmaN}
 \tilde{\sigma}_k &=(-\cos (Qx) \sigma_x+\sin (Qx) \sigma_z,-\sigma_y, \sin (Qx) \sigma_x+\cos (Qx) \sigma_z)
\end{align}
and we have (in the vector representation with respect to spatial direction $i$)
\begin{align}
\Av^{\rm N}_{\rm R} &= -im\alpha_{{\rm R}} (\sigma_y,-\cos (Qx) \sigma_x+\sin (Qx) \sigma_z,0) 
\end{align}
whose Fourier transform is
\begin{align}
\Av^{\rm N}_{{\rm R}}(\qv) &= -im\alpha_{{\rm R}}\delta_{\qv_\perp,0}
\lt[ \delta_{q_x,0} \sigma_y \xvhat - \frac{1}{2}\sum_\pm \delta_{q_x,\pm Q}(\sigma_x \pm i \sigma_z)\yvhat \rt] 
\end{align}
where $\qv_\perp \equiv (0,q_y,q_z)$. 
Uniform component is $\Av^{\rm N}_{{\rm R}}(\qv=0) = -im\alpha_{{\rm R}} \sigma_y \xvhat $.

For the Bloch type spiral, Eq. (\ref{Blochspiral}),  
\begin{align} \label{tildesigmaB}
 \tilde{\sigma}_k &=(-\sigma_x, -\cos (Qx) \sigma_y+\sin (Qx) \sigma_z,\sin (Qx) \sigma_y+\cos (Qx) \sigma_z)
\end{align}
and we have (in the vector representation with respect to spatial direction $i$)
\begin{align}
 A^{\rm B}_{{\rm R},i} &= -im\alpha_{{\rm R}} (\cos (Qx) \sigma_y-\sin (Qx) \sigma_z,-\sigma_x,0) 
\end{align}
and
\begin{align}
\Av^{\rm B}_{{\rm R}}(\qv) &= -im\alpha_{{\rm R}}\delta_{\qv_\perp,0}
\lt[ -\delta_{q_x,0} \sigma_x \yvhat + \frac{1}{2}\sum_\pm \delta_{q_x,\pm Q}(\sigma_y \pm i \sigma_z)\xvhat \rt] 
\end{align}
Uniform component is $\Av^{\rm B}_{{\rm R}}(\qv=0) = im\alpha_{{\rm R}} \sigma_x \yvhat $.
Similar calculations for $\alphav_{\rm R}$ along $\xvhat$ or $\yvhat$ with the use of (\ref{tildesigmaN}) and (\ref{tildesigmaB}) give uniform components shown in Table~\ref{TableSGF}.

For the Weyl type spin-orbit interaction,
\begin{align}
 H_{\rm W} &= -\lambda_{\rm W}\frac{i}{2} \intr c^\dagger (\nablalr\cdot\sigmav) c  
\end{align}
the gauge field is $A_{{\rm R},i} = \lambda_{\rm W}\tilde{\sigma}_i$ and uniform component is as in Table \ref{TableSGF}.

The spin gauge field and Rashba and Weyl spin gauge fields are summarized in Table \ref{TableSGF}.
The Rashba spin gauge field for different directions of $\alphav_{\rm R}$ in the magnetization plane has different coordinate components that are determined by the choice of coordinate system reference point and do not differ physically; the third $\alphav_{\rm R}$ direction gives zero value. Hence the most representative case is $\alphav_{\rm R} || \zvhat$ that is considered in detail above.
Besides, it is seen that $ A_{{\rm s},i}$ and uniform component of the Rahsba gauge field have the same spin polarization direction, i.e., perpendicular to the vector $\mv$ and diagonalization axis $\zvhat$.
The adiabatic components of the conductivity, $\chi_i^{\rm (ad)}$  therefore do not arise from the spin structure and uniform contribution of Rashba gauge field.
The antisymmetric term $\chi_3$ does not arise either.
\begin{table}[hbt]
 \begin{tabular}{c|c|c|c|c|c}
 &  & \multicolumn{3}{c|}{Rashba ($q=0$), $A_{{\rm R},i}$}
 &  \\ \cline{3-5}
&\raisebox{2.6ex}[0cm][0cm]{Spin structure, $A_{{\rm s},i}$}&$\bm{\alpha}_{\rm R} || \hat{x}$&$\bm{\alpha}_{\rm R} || \hat{y}$&$\bm{\alpha}_{\rm R} || \hat{z}$& \raisebox{2.6ex}[0cm][0cm]{Weyl ($q=0$), $A_{{\rm W},i}$} \\
\hline
 N\'eel spiral & $ \frac{Q}{2}\delta_{i,x} \sigma_y $& $im\alpha_{{\rm R}} \sigma_y \delta_{i,z}$ & 0 & $-im\alpha_{{\rm R}} \sigma_y \delta_{i,x}$  &
$- \lambda_{\rm W}\sigma_y \delta_{i,y}$ \\
 Bloch spiral & $ -\frac{Q}{2}\delta_{i,x} \sigma_x $ &  0 & $-im\alpha_{{\rm R}} \sigma_x \delta_{i,z}$ & $im\alpha_{{\rm R}} \sigma_x \delta_{i,y}$  &
$- \lambda_{\rm W}\sigma_x \delta_{i,x}$  
 \end{tabular}
\caption{ Table of uniform components of spin gauge field and Rashba spin gauge fields for N\'eel and Bloch type spirals. \label{TableSGF}}
\end{table}

\begin{figure}
 \includegraphics[width=0.3\hsize]{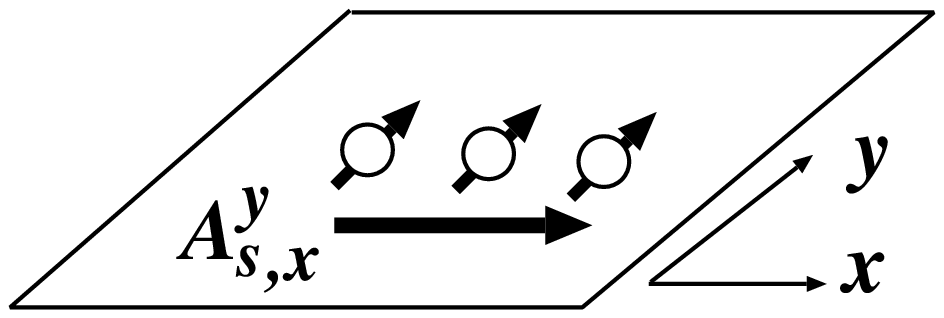}
 \includegraphics[width=0.3\hsize]{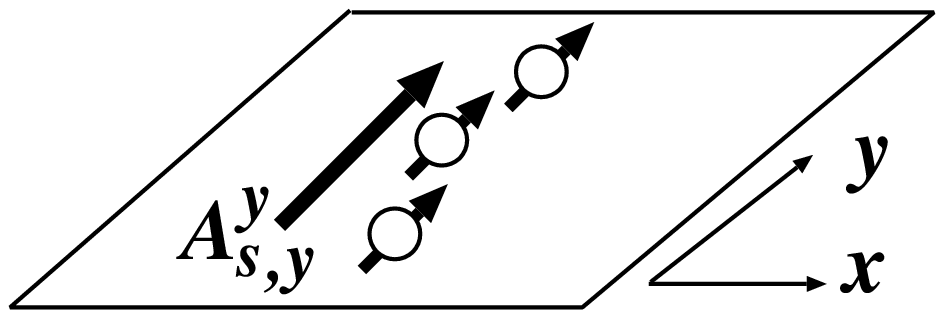}
 \caption{Schematic figure showing spin polarized flow induced by $A_{{\rm s},x}^y$ and 
 $A_{{\rm s},y}^y$. \label{FIGAs}} 
\end{figure}

Having two spin gauge fields from different origins offers interesting possibilities of manipulation of spin and charge. 
A spin gauge field $A_{{\rm s},i}^\alpha$ induces a flow in the direction $i$ polarized along spin direction $\alpha$, namely spin current $j_{{\rm s},i}^\alpha$ (Fig. \ref{FIGAs}).  
Such spin polarized flow does not directly trigger optical responses of material as those responses are governed by charge sector. 
The same spin gauge field maps the spin current to the charge one but the effect on the conductivity tensor is diagonal in simple settings, as the resultant charge flow is along the original direction of spin flow.
Rich possibilities appear if there is another spin gauge field with different symmetry. 
For instance, if we have $A_{{\rm so},j}^\beta$ arising from spin-orbit interaction (like $A_{{\rm R}}$ or $A_{{\rm W}}$), the cross product of the two gauge fields $\sum_\alpha A_{{\rm s},i}^\alpha A_{{\rm so},j}^\alpha$ can induce  off-diagonal charge correlation, $i\neq j$, as a result of conversion of  spin polarization along $\alpha$-direction to charge flow in spatial direction $j$. 
From Table \ref{TableSGF}, we see that such off-diagonal optical response arises for the Weyl spin-orbit interaction with N\'eel spiral and Rashba interaction with Bloch spiral structure. 
Optical response can thus be used to identify spin structures.
Particularly, sudden change of magnetization structures and formation of domains would be detected as emergence of anisotropic and/or off-diagonal optical responses  when external field or temperature is varied.

\section{Directional effects}
As the spin current and the corresponding spin gauge field breaks inversion symmetry, the gauge field appears in the uniform component of the conductivity tensor from the second order. 
The information on the direction of the spin current flow induced by the spin gauge field is therefore smeared in the uniform  optical response considered so far.
Direct effects due to the spin current flow are contained in the directional effects which depend on the wave vector $\qv$ of the external electric field. 
The effects linear in $\qv$ turn out to be linear (or higher-odd order) in the spin gauge field.
Let us briefly study these directional effects.
For the non-uniform component of the conductivity tensor, we need to calculate  ($a,b=\ret,\adv$)
\begin{align}
K_{ij}^{ab}(\qv,\omega,\Omega) &\equiv
\sumkv \tr[v_i(\kv) G^a_{\kv-\frac{\qv}{2},\omega}v_j(\kv) G^b_{\kv+\frac{\qv}{2},\omega+\Omega}]
\end{align}
The Green's function is expanded with respect to $\qv$ as  
\begin{align}
 G_{\kv\mp\frac{\qv}{2},\omega}^a
 &= G_{\kv,\omega}^a  \mp \frac{q_k}{2}  (\partial_{k_k} G_{\kv,\omega}^a)+O(q^2),
\end{align}
to obtain  
\begin{align}
K^{ab}_{ij}(\qv,\omega,\Omega)&=
\sum_k {q_k}  [ K_{ijk}^{ab} -K_{jik}^{ba} ]+K^{ab}_{ij}(\qv=0,\omega,\Omega)
\end{align}
where 
\begin{align}
K^{ab}_{ijk}(\omega,\Omega) & \equiv 
\sumkv \tr[v_i G_{\kv,\omega}^a v_j G_{\kv,\omega+\Omega}^b v_k G_{\kv,\omega+\Omega}^b] 
\end{align}
The trace is calculated in the same way as Eq. (\ref{Kijcalculation2}).
To the first order in the gauge field, the result is ($\omega'=\omega+\Omega$) 
\begin{align}
K^{ab}_{ijk}(\omega,\Omega) 
    & =
\sumkv \frac{2}{ 3\Pi_{k\omega}^a (\Pi_{k,\omega'}^b)^2 }
 \lt[\delta_{ij} (\Av_{{\rm s},k}\cdot\Mv) \lt[k^2 (\pi_{k\omega'}^b)^2+\frac{2}{5}k^4 \pi_{k\omega'}^a  \rt]  \rt. \nnr
 & \lt. 
 +\delta_{ik}(\Av_{{\rm s},j}\cdot\Mv) \lt[  k^2[(\pi_{k\omega'}^b)^2+M^2] + \frac{2}{5}k^4 \pi_{k\omega'}^a  \rt] 
  +\delta_{jk}(\Av_{{\rm s},i}\cdot\Mv) \lt[  k^2[(\pi_{k\omega'}^b)^2+M^2] + \frac{2}{5}k^4 \pi_{k\omega'}^a  \rt]
 \rt]
\end{align}
Here $\Av_{{\rm s}}$ denotes the total spin gauge field including the one due to magnetization structure and spin-orbit interaction. We therefore obtain the conductivity tensor linear (denoted by $^{(1)}$) in both $\qv$ and the spin gauge field as  
\begin{align}
\sigma^{(1)}_{ij}(\qv,\Omega) 
&=\delta_{ij} q_k (\Av_{{\rm s},k}\cdot\Mv) \gamma_1
 + [ q_i(\Av_{{\rm s},j}\cdot\Mv) +q_j(\Av_{{\rm s},i}\cdot\Mv) ] \gamma_2
 \label{sigma1result}
\end{align}
where $\gamma_i$'s are functions of $\Omega$.
Contribution linear in $\qv$ changes sign for opposite light injection, resulting in directional effects like directional dichroism. 
The directional feature arises in the symmetric components in the conductivity to the linear order in the spin gauge field.
As seen from Eq. (\ref{sigma1result}), directional effects arise from the adiabatic component of the gauge field, $\Av_{{\rm s},i}\cdot\Mv $, which vanishes for the spin configurations considered in Table \ref{TableSGF}.
The directional effects predicted by Eq. (\ref{sigma1result}) emerges when $\Av_{\rm s}$ is due to the spin-orbit interaction and when the magnetization $\Mv$ is  uniform.
(Note that Eq. (\ref{sigma1result}) applies to arbitrary direction of  $\Mv$ if $\Mv$ is uniform.)
In fact, directional effect was pointed out in Ref. \cite{Shibata16} for the case of the Rashba spin-orbit interaction treating uniform $M$ perturbatively.

For the Rashba gauge field, Eq. (\ref{Rashbagf}), its uniform component is 
$A_{{\rm R},i}^\alpha(q=0)=im\epsilon_{ij\alpha}\alpha_{{\rm R},j}$ and thus 
the diagonal term of Eq. (\ref{sigma1result}) is proportional to $q_k(\Av_{{\rm R},k}\cdot\Mv)\propto \qv\cdot(\alphav_{\rm R}\times \Mv)$.
The vector $\alphav_{\rm R}\times \Mv$, sometimes called a troidal moment, describes intrinsic velocity of charge as noted in Refs. \cite{Shibata16,Kawaguchi16,TataraReview19}.
For Weyl type, the gauge field is $A_{{\rm W},i}^\alpha =\lambda_{\rm W}\delta_{i\alpha}$, connecting the space and spin diagonally, and so the directional dichroism is with respect to the magnetization direction, $q_k(\Av_{{\rm W},k}\cdot\Mv)\propto \qv\cdot\Mv$. 

\section{Summary}
We have theoretically explored optical properties induced by the second-order effects of spin gauge fields.
The conductivity matrix was calculated in the slowly-varying limit with vanishing  wave vector and angular frequency of the spin gauge field.
Possibility of optical detection of spin structure was pointed out. Additional information is provided by studying the behaviour of optical response under the action of an external magnetic field.
Wave-vector ($q$)-resolved optical response, partially studied here as directional effects, is expected to provide detailed information on the magnetization structures, and this is to be studied in a  future work.

\acknowledgements
This investigation was supported by
a Grant-in-Aid for Exploratory Research (No.16K13853) 
and 
a Grant-in-Aid for Scientific Research (B) (No. 17H02929) from the Japan Society for the Promotion of Science 
and  
a Grant-in-Aid for Scientific Research on Innovative Areas (No.26103006) from The Ministry of Education, Culture, Sports, Science and Technology (MEXT), Japan. 
E.K. would like to thank the Russian Science Foundation (Grant No. 16-12-10340).

\bibliography{/home/gt/References/15,/home/gt/References/gt,3a}
\end{document}